\documentclass{article}

\usepackage{cite}
\usepackage{amsmath,amssymb,amsfonts}
\usepackage{amsthm}
\usepackage{wrapfig}
\usepackage{algorithm}
\usepackage{algpseudocode}
\usepackage{graphicx}
\usepackage{textcomp}
\usepackage{xcolor}
\usepackage{subcaption}
\usepackage{multirow}
\usepackage{booktabs}
\usepackage{soul}
\usepackage{array}
\usepackage{float}
\usepackage{tikz}
\usepackage{longtable}
\usepackage{fancyhdr}
\usepackage{dblfloatfix}
\usepackage{url}
\usepackage[preprint]{neurips_2023}

\usepackage[utf8]{inputenc} 
\usepackage[T1]{fontenc}    
\usepackage{hyperref}       
\usepackage{url}            
\usepackage{booktabs}       
\usepackage{amsfonts}       
\usepackage{nicefrac}       
\usepackage{microtype}      
\usepackage{xcolor}         
\usepackage{array}
\usepackage{caption} 
\title{Foundational Models for Malware Embeddings Using Spatio-Temporal Parallel Convolutional Networks}

\author{Dhruv Nandakumar$^{*}$$^{a}$,
        Devin Quinn$^{a}$,
        Elijah Soba$^{a}$,\\
        Eunyoung Kim$^{a}$,
        Christopher Redino$^{a}$,
        Chris Chan$^{a}$,\\
        Kevin Choi$^{a}$
        Abdul Rahman$^{a}$,
        Edward Bowen$^{a}$\\
        \small $^{a}$Deloitte \& Touche LLP \\
        \small $^{*}$Corresponding author: dnandakumar@deloitte.com \\
}

\begin{document}

\maketitle
\fancyhf{} 
\begin{abstract}
In today's interconnected digital landscape, the proliferation of malware poses a significant threat to the security and stability of computer networks and systems worldwide. As the complexity of malicious tactics, techniques, and procedures (TTPs) continuously grows to evade detection, so does the need for advanced methods capable of capturing and characterizing malware behavior. The current state of the art in malware classification and detection uses task specific objectives; however, this method fails to generalize to other downstream tasks involving the same malware class. In this paper, the authors introduce a novel method that combines convolutional neural networks, standard graph embedding techniques, and a metric learning objective to extract meaningful information from network flow data and create strong embeddings characterizing malware behavior. These embeddings enable the development of highly accurate, efficient, and generalizable machine learning models for tasks such as malware strain classification, zero day threat detection, and closest attack type attribution as demonstrated in this paper. A shift from task specific objectives to strong embeddings will not only allow rapid iteration of cyber-threat detection models, but also allow different modalities to be introduced in the development of these models. 
\end{abstract}

\section{Introduction}
In today's interconnected digital landscape, the proliferation of malware poses a significant threat to the security and stability of computer networks and systems worldwide. As the complexity of malicious tactics, techniques, and procedures (TTPs) continuously grows to evade detection, so does the need for advanced methods capable of capturing and characterizing malware behavior. The current state of the art in malware classification and detection uses task specific objectives; however, this method fails to generalize to other downstream tasks involving the same malware class. It is crucial to create strong embeddings for malware behavior because it enables the development of highly accurate, efficient, and generalizable machine learning models. A shift from task specific objectives to strong embeddings will not only allow rapid iteration of cyber-threat detection models, but also allow different modalities to be introduced in the development of these models. 

In this paper, the authors introduce a novel method for creating strong embeddings of malware behavior using network flow telemetry. The embeddings capture behavioral similarities between similar malware strains while disambiguating distinct strains in the latent space. Furthermore, the authors demonstrate how the pre-trained embedding model can be used for transfer learning to other cyber-threat detection tasks with strong results. The methodology combines convolutional neural networks, standard graph embedding techniques, and a metric learning objective to extract meaningful information from network flows and separate them in the latent space for use in downstream tasks. Overall, the key contributions of the paper are as follows:

\begin{itemize}
    \item A novel method of computing edge weights on connection graphs for graph-based feature learning.
    \item A technique to use Very Sparse Random Projections (FastRP) embeddings (\cite{Chen_Sultan_Tian_Chen_Skiena_2019a}) and graph adjacency matrices to train an embedding model of malware behavior.
    \item Benchmarking performance of the proposed embedding method on three downstream tasks: Malware Classification, Zero Day Threat Detection, and Closest Attack-Type Attribution. 
\end{itemize}

The first few sections of this paper will focus on a literature review of related work, an exploration of the proposed methodology, and an overview of experimental design. The authors will then present findings and experimental results on the performance on the methodology's ability to create strong clusters as well as performance on downstream tasks. The paper will conclude with a summary and discussion of next steps.

\section{Related work}
There have been several past works related to classifying malware using a variety of different methodologies. \cite{kamala} utilized application programming interface (API) based features extracted from sandboxed malware samples to train Decision Tree and Random Forest based models to detect and classify malware strains with good results. \cite{andmal} also utilize tree based techniques to classify malware strains using features generated from static analysis of malware executables. \cite{mamulti} and \cite{meiratten} also propose novel approaches for malware classification using techniques such as system call analysis and attention-based models. 

There has also been research conducted on malware classification using techniques that are conceptually related to the work presented here. \cite{andersonquist} utilized graph representations of malware instruction sequences to identify malware executions. They demonstrated strong classification results but limited their scope to identifying only one malware strain distinctly. Similarly, \cite{DING201873}, \cite{hu_chieuh_shin}, and \cite{kinable} all utilized graph-based feature vectors of system call graphs during malware executions to classify malware strains. In contrast with the work presented here, all of the above approaches use malware execution traces or static analyses to create graphs for classification instead of network flow information. Furthermore, all the methods above are used to train models in a task specific manner and do not lend themselves to transfer learning or other objectives easily.  The methodology proposed in this work, however, involves generating representations of malware behavior that can be used for multiple downstream tasks and is, therefore, conceptually distinct.

Similarly, \cite{roseline}, \cite{alam}, and \cite{WengLo} utilized convolutional neural networks to classify malware strains. However, the input features consisted of malware images which were grey-scale, bit-map representations of malware files. \cite{graphconv} used graph convolutional networks to classify malware strains with graphs constructed on data-flow information. The graphs were constructed to represent data flow within the device on which the malware executed, not across various devices. \cite{zdt1} and \cite{ourzdt} utilized graph features on network flow data for zero day threat detection using autoencoders and metric learning. The approach demonstrated strong performance in identifying anomalies in network flow behavior but only reasonable performance in identifying specific malware types or zero day threats. Furthermore, the approaches utilized were task specific and prone to instability during training. 
 
This paper aims to analyze behavioral differences of malware strains with respect to network connections and information flow across various devices. Our approach captures nuanced information regarding patterns of interaction with command and control servers, etc. which other methods do not. Furthermore, this paper also utilizes a metric learning objective to create embeddings that cluster malware and separate different strains. This allows our model to be more general and extensible to other tasks such as and Zero Day Threat Detection and Malware Classification.

\section{Methodology}

\subsection{Datasets}
The datasets used for our novel architecture were required to contain flow-level information about each event (connection duration, port numbers, timestamp, and number of forward and backward bytes transmitted), the source and destination IP addresses, and attack class labels. Two datasets were used during training and evaluation. The first is Gigas, an organization proprietary dataset consisting of network data over five years. It represents more than 100 real malware sample detonations in our internal malware cyber-range. A subset containing a sample of 20 diverse malware classes and 10,000 examples per class were ordered by timestamp and used as training data.

The second dataset, Malnet, was collected in the form of packet capture files available on malware-traffic-analysis.net (\cite{malnet}). Packet Captures were collected for 16 malware types and converted to flow-level features using the CICFlowmeter tool (\cite{icissp17}). Both the Malnet and Gigas datasets were labelled with class labels referring to the malware strain executed.  Class imbalance was more prevalent in Malnet and served as a benchmark for evaluating model performance in imbalanced scenarios. The per-class support varied from 42,000 to 110 examples per class and had a median of 2,570 examples per class.

\subsection{Feature engineering}

The first novel approach we take is the construction of our directed network-connection graph for network flows captured during malware executions. The nodes on the graph represent distinct Internet Protocol (IP) addresses, and the directed edges represent an aggregation of flow level information between source and destination IP pairs. In this paradigm, duplicate edges can exist between nodes if multiple connections exist between two IP addresses. Each edge is weighted using the connection's network flow attributes as given by the following formula:

\[weight(edge) = \frac{sourceBytes - destinationBytes}{\alpha^{duration}}\]

where $\alpha$  is a hyperparameter, \textit{duration} is the duration of a connection in seconds, and \textit{sourceBytes} and \textit{destinationBytes} represent the amount of information, in bytes, sent from source to destination and vice-versa.  Next, we compute an embedding for each node on the graph using FastRP. FastRP begins by assigning random vectors as embeddings to each node and iteratively averages over a nodes neighbors. The dimension of the embeddings are a tunable hyperparameter and can be modified based on intended downstream use; in this work, we refer to the embedding dimension as $\epsilon$. A node's $\epsilon$-dimensional embedding is a combination of its vector and the average embedding vector of its neighbors. The final embedding of a node \textit{n} is given by: 

\[embedding_n = weight_0 \cdot norm_{l2}(vector_{initial}) + \sum_{i=1}^{n} weight_i \cdot norm_{l2} ({embedding_i} )\]

 where $embedding_i$ is the intermediate embedding of the node given neighbors at the $i^{th}$ degree and $vector_{initial}$ is the initial random vector assigned to the node. It follows that embeddings of nodes seen during inference would be consistent with those generated during training only if the IP behavior in inference data is similar to training data. However, we believe that this assumption is reasonable given the malware behavior is broadly consistent with the objectives of the malware rather than the type of the device on which it executes. The node embeddings will form the basis of the features used in the remainder of this work. 
 
 A key benefit of using FastRP to produce node embeddings is the ability to remove the normalization of raw features. Previously proposed methods rely on the normalization of network flow behavior between several networks to compare them to behavior of malware executions. This methodology can be unreliable, especially when the structure and behavior of the originating networks are vastly different. Our approach eliminates this concern by observing network flow behavior on a per-asset level for a given duration and producing node embeddings for those flows compared to malware behavior. Furthermore, using FastRP removes the need for specific graph feature engineering as seen in the work by \cite{zdt1} and \cite{ourzdt} and allows models to learn from richer behavior-specific embeddings. 

\subsection{ Spatio-temporal example creation}

Once all nodes from malware executions have emeddings, we begin the creation of training and evaluation examples. Malware executions are differentiated not only by the characteristics of individual connection, but also the order in which they occur temporally and spatially (\cite{malbehavior}). That is,  we believe that we can generate richer embeddings of malware behavior if we consider sequences of network flows and the various devices they connect to as opposed to single network flows. Malware execution examples are represented such that they capture the interaction of various nodes on a graph exhibit.

For each malware execution, we first order the network flows in ascending order by timestamp and subset $\beta$ flows. Of the $\beta$ flows, we select the first $\gamma$ IP addresses (either source or destination) seen in the connection. We then construct a ($\gamma$, $\epsilon$ ) dimensional feature matrix, \textit{F}, that contains the FastRP embedding for each selected IP address in the order in which they were seen. \textit{F} now contains feature vectors for all nodes participating in the $\beta$ flows temporally. Next, we construct a binary adjacency matrix for all nodes present in $\gamma$ such that each entry $i,j$ in the matrix will be given by the formula:

\[adjcency(i,j) = \begin{cases}1 & \exists connection_{i,j}\in \beta \\0 & otherwise\end{cases}\]

The adjacency matrix, \textit{A}, represents spatial connections between interacting nodes on a graph and has dimension $(\gamma, \gamma)$. \textit{A} and \textit{F} form the input features for one example. Similarly, a sliding window of width $\beta$ can be used to create several examples per malware execution for all malware executions with a corresponding label.

\subsection{Model architecture}
\begin{figure}[h!]
    \centering
    \includegraphics[width=.60\linewidth]{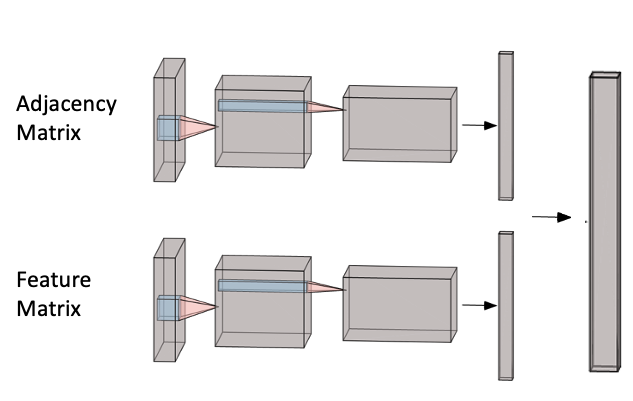}
    \caption{Parallel Convolutional Model Architecture}
    \label{fig:model_arch}
\end{figure}

We propose a spatio-temporal parallel convolutional network (ST-PCN) architecture that processes the spatial and temporal aspects of malware execution graphs in parallel for the creation of strong embeddings. Our model architecture begins with two sets of convolutional layers that convolve \textit{A} and \textit{F} independently to produce a 32 dimensional vector embedding for each matrix. Both vector embeddings are then concatenated to produce a final spatio-temporal embedding which is passed into the metric learning objective function. The model architecture is depicted in Fig. \ref{fig:model_arch}. 

\subsection{Training and evaluation}
The ST-PCN architecture is trained using a metric learning objective to maximize inter-class separation in the embedding space while simultaneously minimizing intra-class separation. Particularly, we utilize a softmax-based additive angular margin loss (\cite{arcface}) as our loss function which is backpropagated through the entire ST-PCN model and is computed on the concatenated embedding and corresponding malware label. 


Furthermore, we will introduce a 'holdout' malware class which will be held out from training and validation data entirely when training the ST-PCN models and evaluating them on downstream tasks. Downstream task evaluation will also assess the performance of our models on holdout data to estimate the performance of the approach on novel malware. This is particularly helpful when evaluating the efficacy of the ST-PCN on a zero day threat detection task. All experiments conducted below require a compute instance with at least 64 gigabytes of Random Access Memory ( RAM ) and a 32 core processor.

\section{Experimental design}

Training and testing are performed on a random 70/30 train-test split of input matrices. Given that the objective of the ST-PCN is to produce strong embeddings of malware behavior for downstream tasks, the primary evaluation will be focused on the embedding clusters and associate metrics such as silhouette scores, Rand indices, and cluster completeness and homogeneity scores. Furthermore, performance of the embeddings will also be evaluated based on the performance of complex downstream tasks.  All downstream tasks will use pre-trained embeddings from the ST-PCN with no fine-tuning.

\subsection{Embedding by attack class}

The metric-learning objective involves training an ST-PCN model on a metric learning objective that aims to separate malware by their respective strain. Compared to recent work (\cite{ourzdt}) which uses metric learning on cyber attack types such as botnet or ransomware attacks, this work uses more granular malware strain labels because each malware strain exhibits distinct behavior that could be used for multiple attack campaigns. For example, a Bazaloader trojan could be used in an overall attack campaign to deliver ransomware, keystroke loggers, or any other malware. We believe this will lead to better separability in the latent space. As mentioned in the Datasets section, we use malware labels from two datasets consisting of a total 36 malware classes. 

The produced FastRP embeddings only consider a node's neighbors up to the second degree. Consequently, each FastRP embedding is a weighted sum of three intermediate embeddings: the node's initial vector, it's vector relative to it's neighbors, and it's vector relative to it's neighbors' neighbors. The weights of each vector are 1, 0.5, and 0.5, respectively. Furthermore, all experiments and results set $\alpha$ as 1.15, $\beta$ as 128, and $\gamma$ and $\epsilon$ as 32.

\subsection{Malware classification}

This downstream task evaluates the performance of pre-trained ST-PCN embeddings on classifying malware types. Evaluation is conducted on embeddings produced for the test-set of malware classes and the holdout malware class. In the malware classification experiments that follow, we utilize a Random Forest Classifier fit on the training set of embeddings.

\subsection{Zero day threat detection}

In this task, we evaluate the models' performance on differentiating a malware class seen during training to a holdout class. In order to estimate the performance of ST-PCN embeddings on a zero day threat detection task, we utilize a Euclidean distance-weighted K-Neighbor classification model fit on the training set of embeddings with the number of neighbors set to 350. 

 During evaluation, we compute maximum class membership probability for each test and holdout example on the training classes. The complement of this probability is then computed to represent the probability of an example being a Zero Day Threat, referred to henceforth as the ZDT probability. The ZDT probabilities are then thresholded to compute metrics. Given the class imbalance between test examples and holdouts, we utilize precision, recall, and area under the precision-recall curve as evaluation metrics.

\subsection{Closest attack type attribution}

 The objective of this downstream task is to determine, for any given holdout example, what the most similar malware seen during training to it is. This allows us to evaluate the fidelity of the organization of the latent space and examine if similar malware types are truly located close to each other. 

 The same K-Neighbors classification used in the Zero Day Threat detection task is used to compute the closest attack type and probability for every holdout example. Then, for each holdout class, we extract the two most common classifications and compute the average probability for each classification.

\section{Results and discussion}

\subsection{Cluster and embedding analysis}

\begin{figure}[h!]
    \centering
    \begin{subfigure}[b]{0.45\textwidth}
        \includegraphics[width=\textwidth]{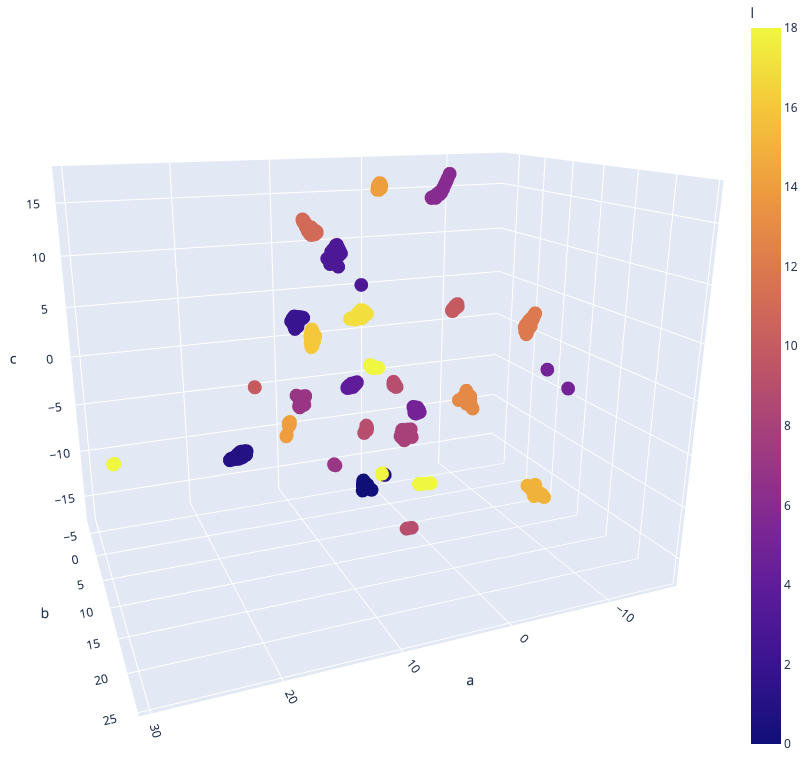}
        \caption{Gigas dataset embeddings}
        \label{fig:gigas_space}
    \end{subfigure}
    \hfill
    \begin{subfigure}[b]{0.45\textwidth}
        \includegraphics[width=\textwidth]{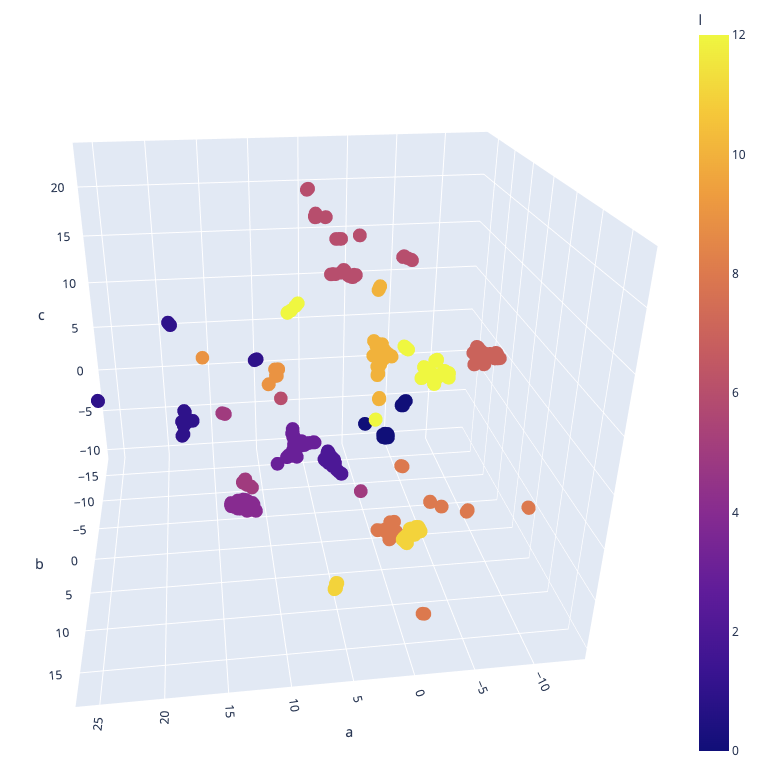}
        \caption{Malnet dataset embeddings}
        \label{fig:malnet_space}
    \end{subfigure}
    \caption{Dataset embeddings}
    \label{fig:dataset_embeddings}
\end{figure}

In order to visualize and evaluate the embeddings produced by the ST-PCN model, we produce embeddings for all example matrices in the test-set. Once embeddings are computed, we utilize Uniform Manifold Approximation and Projection (UMAP) (\cite{mcinnes_healy_melville_2020}) to produce 3-dimensional representations of the embeddings for visualization. Visualizations for both Gigas and Malnet datasets are shown in Fig \ref{fig:gigas_space} and \ref{fig:malnet_space}, respectively. In both figures, the points are colored by malware class.

From the figures, we see that the ST-PCN produces well separated clusters for malware classes in both datasets. We also see that a few malware strains tend to have multiple tight clusters as opposed to a single larger cluster. We believe that this is due to sub-variations in malware behavior based on external factors such as the operating system on which they execute. In general, however, malware classes are well separated into either tight clusters or slightly diffuse cluster clouds. This analysis is also supported by strong evaluation metrics presented in Table \ref{table:1}. 

\begin{table} [H]
\centering
\caption{ST-PCN embedding metrics}
\begin{tabular}{cccccc} 
\toprule
Dataset & Silhouette & Completeness & Homogeneity & Rand \\
\midrule
Gigas & 0.69 & 0.86 & 0.94 & 0.97 \\
Malnet & 0.66 & 0.71 & 0.97  & 0.96 \\
\bottomrule
\end{tabular}
\label{table:1}
\end{table}

\subsection{Malware classification}

Performance of the ST-PCN model on the classification task showed consistently strong performance on complete test sets of both Gigas and Malnet datasets with 19 and 13 classes respectively. Macro and Lowest class-specific precision, recall, and AUC scores are provided in Tables \ref{table:2} and \ref{table:3}. 

\begin{figure}[H]
    \centering
    \begin{minipage}[b]{0.45\textwidth}
        \centering
        \captionof{table}{Malware classification metrics on Gigas test data}
        \begin{tabular}{cccc} 
         \toprule
        Value & AUC & Precision & Recall \\
         \midrule
         Macro Avg. & 0.99 & 0.99 & 0.99 \\
         Minimum & 0.98 & 0.97 & 0.97 \\
         \bottomrule
        \end{tabular}
        \label{table:2}
    \end{minipage}
    \hfill
    \begin{minipage}[b]{0.45\textwidth}
        \centering
        \captionof{table}{Malware classification metrics on Malnet test data}
        \begin{tabular}{cccc} 
         \toprule
        Value & AUC & Precision & Recall \\
         \midrule
         Macro Avg. & 0.99 & 0.99 & 0.99 \\
         Minimum & 0.98 & 0.94 & 0.97 \\
         \bottomrule
        \end{tabular}
        \label{table:3}
    \end{minipage}
\end{figure}

In order to test performance of the ST-PCN when a novel class is present in training data, we re-ran the above experiment with introducing a distinct holdout class in the embedding data and training data for the classification model 5 times. Aggregate results show good performance on both datasets but with a noticeable drop in performance in tables \ref{table:4} and \ref{table:5}. However, overall metrics still indicate that the ST-PCN can generalize with reasonably strong performance.

\begin{figure}[H]
    \centering
    \begin{minipage}[b]{0.45\textwidth}
        \centering
        \captionof{table}{Malware classification metrics on Gigas test data with holdout}
        \begin{tabular}{cccc} 
         \toprule
        Value & AUC & Precision & Recall  \\
         \midrule
         Macro Avg. & 0.95 & 0.97 & 0.91 \\
         Minimum & 0.40 & 0.32 & 0.15 \\
         \bottomrule
        \end{tabular}
        \label{table:4}
    \end{minipage}
    \hfill
    \begin{minipage}[b]{0.45\textwidth}
        \centering
        \captionof{table}{Malware classification metrics on Malnet test data with holdout}
        \begin{tabular}{cccc} 
         \toprule
        Value & AUC & Precision & Recall  \\
         \midrule
         Macro Avg. & 0.93 & 0.93 & 0.90 \\
         Minimum & 0.31 & 0.19 & 0.25 \\
         \bottomrule
        \end{tabular}
        \label{table:5}
    \end{minipage}
\end{figure}

\subsection{Zero day threat detection}

The Zero Day Threat Detection task was carried out and evaluated for 5 different holdouts per dataset. Overall, the ST-PCN exhibited consistently strong performance on the Gigas dataset. Performance of the model on the Malnet dataset was also generally strong, but some classes had lower performance. This is ostensibly due to the significantly lower number of training examples per class in the Malnet dataset compared to Gigas. All results are shown in Table \ref{table:6} in the \textit{AUC}, \textit{Precision}, and \textit{Recall} columns. Furthermore, we already see strong performance improvements compared to recent work by \cite{zdt1} and \cite{ourzdt} with no fine tuning and a relatively simple classification model. 

\begin{table} [H]
\centering
\caption{Zero Day Threat detection and CATA metrics}
\begin{tabular}{ccccc | cc} 
 \toprule
Dataset & Holdout & AUC & Precision & Recall & CATA & Probability  \\
\midrule
Gigas & Nanocore & 0.90 & 0.90 & 0.85 & Meterpreter & 91\% \\
Gigas & Azorult & 0.96 & 0.99 & 0.91 & Ursnif & 88\%\\
Gigas & Ursnif & 0.89 & 0.98 & 0.72 & Nymeria & 91\% \\
Gigas & Trickbot & 0.96 & 0.98 & 0.92 & Xtremerat & 93\% \\
Gigas & Lokibot & 0.95 & 0.99 & 0.92 & Netwire & 88\%\\
Malnet & Bazaloader & 0.97 & 0.96 & 0.94 & SquirrelWaffle & 81\%\\
Malnet & Astaroth & 0.82 & 0.95 & 0.70 & Hancitor & 96\% \\
Malnet & Mantabuchus & 0.68 & 0.76 & 0.65 & Monsterlibra & 76\% \\
Malnet & Valak & 0.92 & 0.94 & 0.91 & Hancitor & 90\% \\
Malnet & Qakbot & 0.88 & 0.93 & 0.84 & Gozi & 89\% \\
\bottomrule
\textbf{Average} && \textbf{0.89}& \textbf{0.94} & \textbf{0.83} \\
\hline
\end{tabular}
\label{table:6}
\end{table}

\subsection{Closest attack type attribution (CATA)}

The CATA for every holdout evaluated in the Zero Day Threat Detection section is provided in Table \ref{table:6} in the \textit{CATA} and \textit{Probability} columns. For each holdout, the closest attack type  in the training set is provided alongside a probability score.

Analyzing some results above reveals that the embedding space produced by the ST-PCN is, in fact, well-ordered by behavior. For example, we see that the downstream model indicated that the \textit{Nanocore} holdout is closest to the \textit{Meterpreter} malware. External analysis indicates that both malware types are Remote Access Trojans (RATs) and therefore exhibit similar behavior. We see similar results for 
\begin{itemize}
    \item \textit{Bazaloader} and \textit{SquirrelWaffle} which are both 'malware downloaders' which perform the function of downloading malware execution suites from remote sources.
    \item \textit{Astaroth} and \textit{Hancitor} which are also software packages used to facilitate dropping or downloading of malware.
    \item \textit{Qakbot} and \textit{Gozi} which are malware strains used for information stealing and keylogging as part of larger attack campaigns.
\end{itemize}

We believe that enriching model results with CATA attributions allow for enhanced model interpretability as well as increased utility of model alerts to end users such as Threat Hunters who could use model outputs for further investigation in an network.

\section{Conclusion, Limitations, and Future Work}

In this work, we have presented a novel, foundational method for producing embeddings of network-agnostic malware behavior for downstream use. In the proposed methodology, we utilize graph-based embeddings of asset behavior as a feature engineering step to a ST-PCN. We also demonstrate the strong performance of the embedding model both independently and using historically non-trivial tasks such as Zero Day Threat Detection and CATA. We believe our results show the potential to prevent cyber-operator fatigue and allow rapid iteration of malware-specific machine learning techniques.

While the metrics on downstream tasks indicate strong results, we hypothesize that more complex, task specific architectures can achieve even better performance. The out of the box methods used in our experiments primarily leverage the well organized latent space to perform their respective tasks. However, task specific architectures can further manipulate the latent space and extract meaningful features in a way that could not be done with our reported methodologies. In tasks such as Zero Day Threat Detection, even a small increase in performance can have a big impact. 

Furthermore, we also hypothesize that incorporating multi-dimensional or multi-modal raw feature data will further strengthen the ST-PCN model. Data such as malware severity scores or endpoint based logs can provide context about malware executions beyond network flow patterns. Incorporation of this type of data could not only alleviate the strict dependence on network flow data, but also introduce new downstream tasks our model can generalize to. 

A limitation to this work is the limited availability of malware strains present in our training and validation data. Although we made a concerted effort to capture a wide diversity of malware, the set of known and unknown malware is larger than the data available to us. Adding more strains has the potential to make our embeddings more general but could impact performance of the model. Consequently, further training and validation on wider sets of malware data is an important next step in the development of this framework. 

Future work will aim to improve the ST-PCN architecture and performance by incorporating multi-modal data and introducing more complex downstream task models. In addition, we will attempt to use more malware strains and investigate how it affects our performance across downstream tasks. If performance maintains or exceeds what is reported, it will indicate that our model is a good candidate for generalizing malware in cyber specific machine learning applications.

\bibliographystyle{plainnat}
\bibliography{ref}

\end{document}